# A superconducting nanowire spiking element for neural networks

E. Toomey[1], K. Segall[2], M. Castellani[1], M. Colangelo[1], N. Lynch[1], and K. K. Berggren[1]

[1]Massachusetts Institute of Technology, Department of Electrical Engineering and Computer Science, Cambridge, MA 02139, USA

[2]Colgate University, Department of Physics and Astronomy, Hamilton, NY 13346, USA

**ABSTRACT**

As the limits of traditional von Neumann computing come into view, the brain's ability to communicate vast quantities of information using low-power spikes has become an increasing source of inspiration for alternative architectures. Key to the success of these largescale neural networks is a power-efficient spiking element that is scalable and easily interfaced with traditional control electronics. In this work, we present a spiking element fabricated from superconducting nanowires that has pulse energies on the order of ~10 aJ. We demonstrate that the device reproduces essential characteristics of biological neurons, such as a refractory period and a firing threshold. Through simulations using experimentally measured device parameters, we show how nanowire-based networks may be used for inference in image recognition, and that the probabilistic nature of nanowire switching may be exploited for modeling biological processes and for applications that rely on stochasticity.

**Keywords:** *spiking neural networks (SNNs), superconducting nanowire, neuromorphic computing, spiking hardware*

## 1. INTRODUCTION

Recent years have witnessed the growth of brain-inspired computing architectures in response to the stagnation in performance of traditional systems[1], offering the opportunity to advance both electronics and our understanding of how the brain operates. Of the existing neuromorphic architectures, spiking neural networks (SNNs) are among the most bio-realistic approaches, relying on electrical spikes analogous to action potentials in order to compute with high energy efficiency and speed[2].

At the heart of SNNs are devices or simple circuits that serve as a spiking element or "soma", generating electrical spikes with varying degrees of bio-realism while maintaining low power. To-date, spiking behavior has been explored in a variety of hardware platforms, including CMOS, magnetic materials, and Mott insulators, but each of these is accompanied by certain drawbacks. For example, SNNs made from CMOS[3] allow for easy integration with external circuitry, but usually require many components to generate spiking and are not as energy-efficient as the human brain. On the other hand, magnetic materials that generate spiking by harnessing the spin-torque effect[4] suffer from small on/off ratios that lead to low signal levels[5]. Relaxation oscillators using Mott insulators[6] can also generate spiking, but maintain slow time constants and high pulse energies. These shortcomings motivate the need for a robust, scalable, and power-efficient device that naturally generates spiking and integrates easily with existing control circuitry.

Superconductors are prime candidates for spiking applications due to their negligible static power dissipation and rapid switching speeds. Building off of a scheme first implemented in Josephson junctions[7], we recently proposed a nanowire-based artificial neuron[8] whose soma generates pulses by taking advantage of coupled relaxation oscillations[9] that occur in nanowires as they switch between the superconducting and normal states. In addition to having low switching dissipation, superconducting nanowires can be densely packed for scaling and have high output voltages[10] that enable compatibility with external CMOS control circuitry[11], making them appealing candidates for the development of a largescale neural network that can interface with traditional systems. Here we present a low-power nanowire spiking element based on our previously proposed design that has a pulse energy on the order of ~10 aJ. We experimentally demonstrate that the device reproduces several bio-realistic behaviors and use measured characteristics to simulate two potential applications of a nanowire-based neural network.

## 2. SOMA EXPERIMENTS

Figure 1a shows a simplified circuit schematic of our device, based on our previously reported design[8]. Two nanowire relaxation oscillators are linked together in a superconducting loop and act analogously to the sodium and potassium ion channels in a simplified action potential model[12,7]. To operate the device, a bias current from the top port biases both oscillators right below their critical currents $I_c$, or the point at which they transition into the resistive state. When an input current is applied from the left, the bias and input sum together to trigger the "main oscillator," which fires and adds flux into the loop in the form of a circulating current, similar to the influx of Na+ in a biological neuron. This additional current then fires the "control oscillator", which acts analogously to the K+ outflux by removing current from the loop, resetting the soma and allowing it to fire again. More details about the operating principles of this device are described in our earlier work that simulated its behavior[8].

Figure 1b shows scanning electron micrographs of a nanowire soma fabricated from thick (~25 nm) niobium nitride. Each of the two relaxation oscillators consists of a 60-nm-wide nanowire switching element placed in parallel with a shunt resistor $R$. In order to support relaxation oscillations, meandered nanowire inductors of magnitude $L$ were patterned in series with the switching element to give an $L/R$ time constant on the order of nanoseconds. Details of the fabrication process and design may be found in the Supplemental Material.

To gauge the spiking characteristics of the soma, we first measured its frequency response when driven as a single oscillator by overbiasing it from the bias port without applying an input pulse. In this operation, the two oscillators fire simultaneously, since there is no input signal to induce a phase shift between them. Figure 1c shows the oscillation frequency of the overbiased soma without the application of an input pulse, revealing that the soma starts operating as a single oscillator around ~82 $\mu$A. This frequency response can be well-explained by our LTspice model[13] for a nanowire soma, using circuit parameters measured from an isolated oscillator that was patterned alongside the device (see Supplemental Material Figure S1). The red data points in Fig.1c show the simulated response of the nanowire soma when $I_c = 38.5$ $\mu$A, $L = 6$ nH, $R = 1.2$ $\Omega$, and the loop inductors are both 12 nH. By comparing both the frequency response as well as the time-domain characteristics of the experimental and simulated results, we can conclude that the LTspice soma model sufficiently reproduces real dynamics of the physical device.

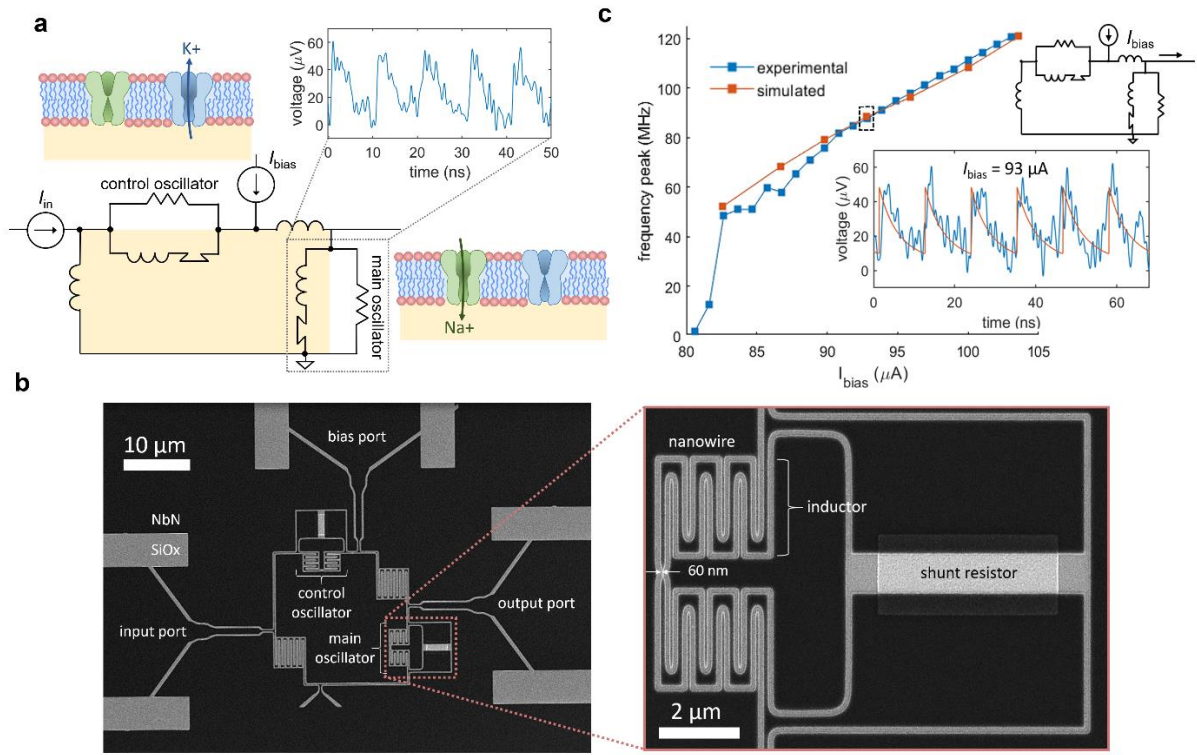

**Figure 1: Device design and characterization.** (a) Simplified circuit schematic of the neuron, consisting of two superconducting nanowire relaxation oscillators linked together in a superconducting loop.[8] The main oscillator adds circulating current into the loop, acting like the Na+ influx in a biological neuron. The control oscillator removes that current, similar to the K+ outflux, resetting the cell and allowing it to spike again. Inset shows experimentally measured oscillations from an isolated relaxation oscillator. (b) Scanning electron micrograph of a fabricated soma. Dark areas are the niobium nitride film, while the grey outlines are the underlying substrate. Inset: Scanning electron micrograph showing an enlarged view of one of the relaxation oscillators. (c) Frequency response of the soma when it is driven as a single oscillator by overbiasing the device without applying an input pulse. Inset shows the time domain at $I_{bias}$ = 93 µA, comparing the experimental results (blue) to the LTspice simulation (red).

With the updated LTspice model, we can experimentally test the soma's spiking operation and use simulations to better understand its behavior. Figure 2a-b show the outputs of the measured and simulated soma in response to an input pulse when the bias current is 76.3 $\mu$A, below the point at which the soma acts as a single oscillator. As demonstrated in the figure, the soma only spikes in response to the input pulse, in agreement with the expected operation. To ensure that the spikes are coming from the phase-shifted firing of both oscillators, we also examined the voltage signals of the input port and the bias port, and compared them to the simulated responses (see Supplemental Material Figure S2).

Spiking reproducibility was assessed by gathering statistics on the interspike interval $\Delta t$. Figure 2c shows a histogram of the interspike intervals from 100 captured waveforms, resulting in a distribution with a mean of 50.4 ns and a standard deviation of 6.46 ns. This spread is comparable to what has been observed in human motoneurons, where the standard deviation is 5-10% of the mean interspike interval[14], suggesting that the nanowire soma possesses relative bio-realistic timing characteristics despite firing at a much faster frequency (~10 MHz vs. ~10 Hz).

A key feature of biological neurons is the existence of a firing threshold, or a minimum input signal required to initiate spiking for a given resting potential[15]. In the nanowire soma, the resting potential is dictated by the bias current—a larger bias current raises the resting potential and decreases the firing threshold. Figure 2d shows the threshold response of the fabricated soma, measured as the mean voltage output of 500 sequential traces for a given input current. This measurement translates into firing probability, since the mean voltage of the 500 measurements will be higher if the soma spikes more often. Comparing the mean voltage output of the different bias curves at the same input level shows that the firing probability increases with increasing bias current, as expected.

The curves in Fig. 2d also reveal that the nanowire soma has an S-shaped firing probability as a function of input current. This trend is attributed to the stochastic nature of nanowire switching in real measurements, where thermal and quantum fluctuations cause premature switching at currents below the critical current[16], leading to a switching probability that increases with the total applied bias. For low bias currents, we observed that the firing probability decreases at some point as the input current increases. One possible explanation for this phenomenon is if the inductances of the two oscillators are slightly unequal due to material inhomogeneities or other defects, leading to differences in the amount of flux they contribute to the loop (see Supplemental Material Section V for a more detailed discussion). Choosing a higher bias current without this signature allows for optimal operation of the device where there is a large difference in firing probabilities between the low and high input current levels.

While stochasticity like the variable firing threshold observed here is usually avoided in electrical systems, biological neurons have firing probabilities that have inspired many neuromorphic applications that take advantage of probabilistic switching, making nanowires a logical hardware platform for these types of applications. An example of harnessing stochastic firing for a neuromorphic application is demonstrated later in this paper.

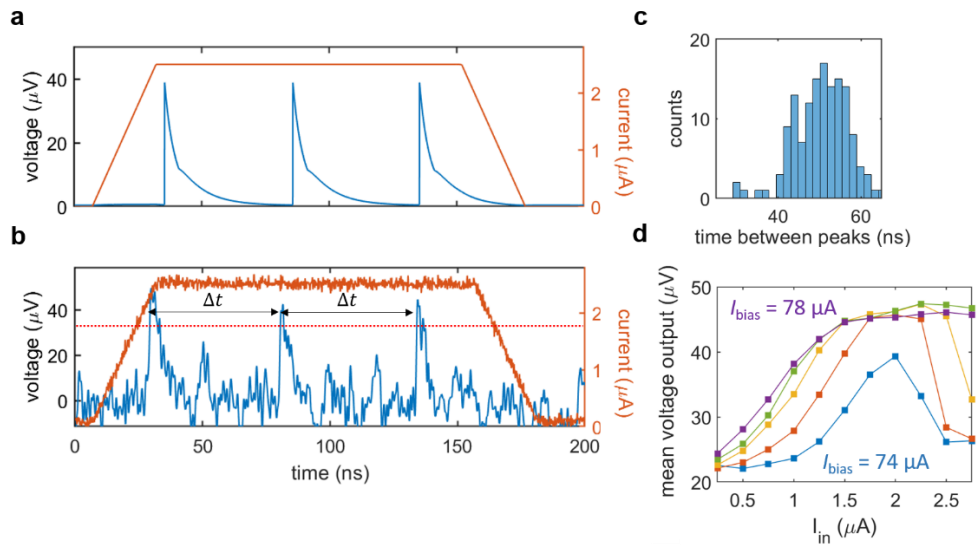

**Figure 2: Response to an input pulse.** (a) Simulated spiking of the nanowire soma when biased at 76.3 µA. Orange trace is the input pulse while the blue trace is the output. (b) Spiking measured from the fabricated soma under identical conditions. The red dashed line represents the threshold used to measure the presence of a spike, while $\Delta t$ indicated the interspike interval. (c) Histogram of the interspike intervals measured from 100 sequential waveforms. (d) Firing probability as a function of input current, measured as the mean voltage output of 500 sequential traces. The different curves represent different bias currents, swept from 74 µA to 78 µA in 1 µA increments. Input pulses had a 150 ns pulse width and 20 ns edge times.

In addition to a threshold response, biological nanowires display a refractory period[15], or a required "resting period" between two inputs so that both elicit their own output spikes. Figure 3 shows the measured and simulated output of the soma in response to two identical input pulses that were gradually brought closer together. In Fig. 3a-b, the time between input pulses was sufficient for both inputs to generate separate output spikes. In Fig. 3c, however, the soma only spikes once. The simulation reveals that the second input pulse ends right as the first spike relaxes, suggesting a refractory limit in which the main oscillator is not sufficiently biased during the second input pulse to fire again. Figure 3d shows histograms of 200 repeated measurements of the time of each output spike as the input pulses move closer together. The gradual collapse of two distinct histograms into one verifies reproducibility of the refractory period observed in the individual traces.

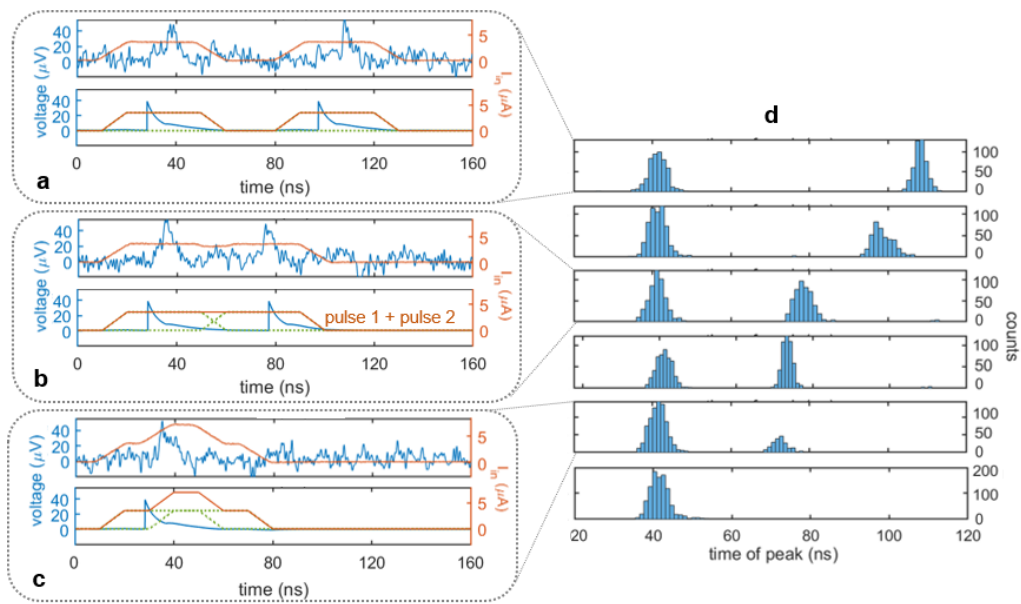

**Figure 3: Demonstration of the refractory period.** (a) Output voltage (blue) when rising edges of the input current pulses (orange) are separated by ~70 ns. Top panel is the experimental waveforms while the bottom panel is the simulation. For both cases, $I_{bias}$ = 75.9 $\mu$A. (b) Same as (a), but with the rising edges separated by ~ 40 ns. The orange trace is the sum of the two input pulses, whose rising and falling edges are indicated by the green dashed lines. (c) Same as (b), but with the rising edges separated by ~ 20 ns. The presence of only one output pulse indicates the effect of the refractory period. (d) Histograms represent 200 measurements of the time at which a spike occurs as the input pulses are brought closer together. The two distinct histograms collapse into one as the refractory period is approached.

## 3. APPLICATIONS

Using the LTspice model of the nanowire soma updated with experimentally measured device parameters, we can simulate how a nanowire-based network may be used in an inference chip for applications like image classification. Figure 4a shows a simple 3×3 image of the letter "z" that is part of a test set originally developed for physical memristor circuits[17] and recently simulated in Josephson junction neural networks[18]. The complete set consists of three "ideal" images of the letters "z", "v", and "n", as well as nine single-pixel-error images per letter (see Supplemental Material Figure S5). To identify these images, we can use a 9×3 neural network consisting of nine input pixel neurons and three output letter neurons. The pixel colors determine the input current to each pixel, with grey pixels corresponding to an input current of 4.6 $\mu$A and white pixels corresponding to an input current of 0 $\mu$A.

The output of each pixel neuron is fed into the input of each letter neuron using an inductive synapse with inductive coupling. As described in our previous work[8], the weight of each connection maps to the magnitude of the synapse inductor, with higher weights represented as lower inductances, leading to more synaptic current. Following methods previously applied to Josephson junction networks[18], we used a basic neural network script[19] written in Python to solve for the weights, which determined the synapse inductances in our LTspice circuit. Due to the limited size of the data set, we used all 30 images for both training and testing, leading to 100% classification. An example of the network's performance when fewer images were used during training may be found in the Supplemental Material (see Figure S6).

Figure 4b shows the pixel input currents and the letter output voltages for all 30 images. As shown by the network's output, each letter neuron fires 10 times, indicating correct classification of both the ideal and error images, which is to be expected since all of the images were used during training. Although the nanowire soma does not yet have a scheme for unsupervised learning, these results demonstrate that it may serve as an energy-efficient inference platform in a hardware neural network designed to perform a specific task.

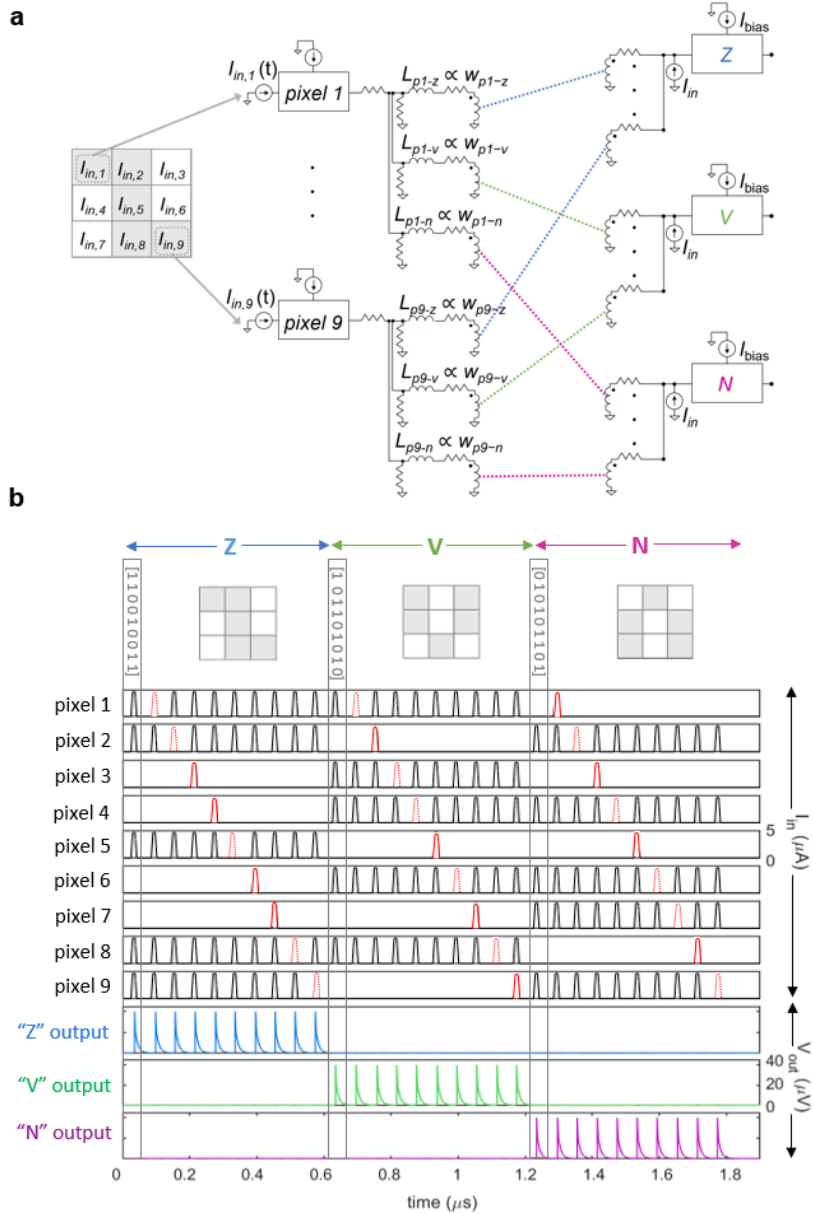

**Figure 4: Recognition of 3×3 pixel images in a simulated network.** (a) Simplified circuit schematic of the network, with the ideal "z" image shown as an example. (b) Simulation of each of the 30 test images in the network. The first nine panels show the input current sent to each pixel neuron, with missing bits represented by red dashed lines and extra bits represented by solid red lines. The three panels shows the output voltage of each of the three letter neurons. Grey boxes indicate the "ideal" image for each letter, followed by their nine single-pixel error images. For every synapse, the inductance was scaled relative to 0.1 µH, and the series resistor was 5 Ω and the parallel resistor was 25 Ω.

Although the image recognition example above used a soma model that fires deterministically, or exactly when the current through the nanowire exceeds its $I_c$, the threshold measurements in Fig. 2d showed that real superconducting nanowires have a firing threshold that varies stochastically due to the effects of noise and fluctuations.

As mentioned previously, biological neurons also have firing probabilities, which has inspired theories of how stochastic behavior plays a critical role in the brain's operation. One key example of harnessing stochasticity is the winner-takes-all (WTA) theory[20], which suggests that the brain develops selectivity through competition between excitatory neurons that share a set of inhibitory connections. WTA subcircuits are suspected to be repeated throughout the brain, creating selective responses that together dictate a cumulative behavior. This functionality has inspired the use of WTA in artificial networks for filtering, image recognition, and decision-making[21].

Figure 5a shows a schematic of a two-inhibitor WTA network presented by Lynch et al.[22] When a set of input neurons $X_{1:n}$ fires, they trigger a set of output neurons $Y_{1:n}$ that have stochastically varying thresholds. Competition between output neurons is facilitated by two inhibiting neurons $Z_s$, the stability inhibitor, and $Z_c$, the convergence inhibitor. $Z_s$ is biased so that it fires when at least one output neuron is firing, whereas $Z_c$ is biased so that it only fires if two or more output neurons are firing. $Z_c$ eventually forces all but one neuron to stop firing, while $Z_s$ continues to fire in order to stabilize the network and suppress all but the dominant neuron. Since $Y_{1:n}$ are identical and stochastic, they have equal probability of winning if all inputs are active.

We can take advantage of the nanowire's intrinsic stochasticity in order to simulate a WTA competition. We amended our soma model to include Gaussian white noise sources, keeping the noise bandwidth constant at 1 GHz and varying the noise amplitude until the firing probability was comparable to what we observed experimentally. Figure 5b shows relative agreement between the simulated firing probability and the experimentally measured probability converted from the mean voltage curves of Fig. 2d, suggesting that our model is suitable for implementing a realistic firing rate. With these adjustments to our model, we designed a simple two-inhibitor WTA network with three deterministic inputs and three stochastic outputs. As in the pattern recognition network, neurons were connected via inductive coupling.

Figure 5c displays the outcomes of 100 repeated competitions. Each output neuron won roughly the same number of times (20-27), indicating that the winner is selected through probability. Figure 5d shows the time domain of a competition when $Y_3$ was the winner. As seen by the output voltages from each neuron, $Z_c$ only fires as long as two outputs are active, while $Z_s$ continues to fire along with the winner, keeping the other outputs suppressed. 29 of the 100 competitions resulted in no winner, usually with all three output neurons being shut off. In some cases, this seemed to be due to continued firing of $Z_s$, despite the outputs being turned off, which could be caused by instability resulting from biasing it close to its critical current. This issue may be avoided by optimizing our circuit parameters to increase the synaptic currents so that the biases to the neurons can be reduced.

Despite the competitions that resulted in no winner, the similar outcomes of the three output neurons illustrate how we can build a network that uses probabilistic firing to produce a meaningful output from a set of inputs, and that superconducting nanowires have enough inherent stochasticity to support this type of competition.

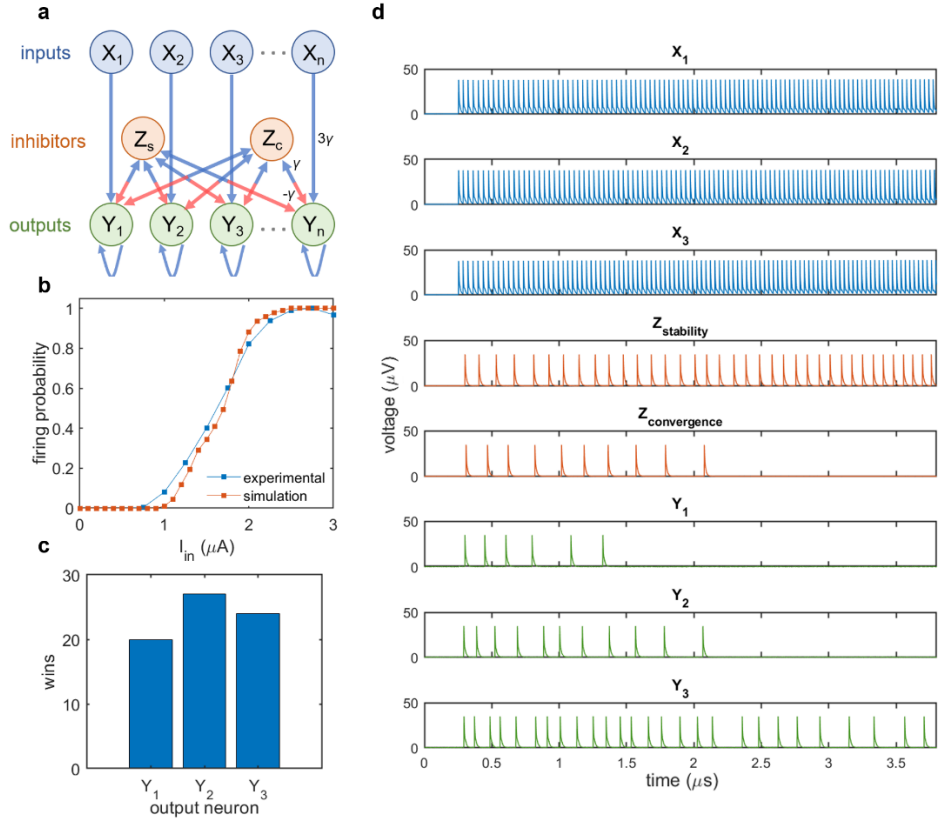

**Figure 5: Winner-takes-all competition in a simulated network.** (a) Basic schematic of a two-inhibitor WTA network. Excitatory connections are shown in blue and inhibitory connections are shown in red, with their weights defined relative to the dimensionless parameter $\gamma$. (b) Experimental and simulated firing probabilities. The simulated probability was calculated by recording the number of times the soma fired out of 50 trials for each input current level. The experimental bias was 74.5 $\mu A$, while the simulated bias was 76.6 $\mu A$ with a noise amplitude of 800 nA. (c) Results from 100 repeated WTA competitions. Each neuron won between 20-27 competitions, while 29 of the trials had no winner. (d) Example of output voltages during a single WTA competition. In this case, $Y_3$ wins. For this simulation, the bias of the output neurons was 76.4 $\mu A$ and synapse inductors were scaled relative to 0.77 $\mu H$, and the couplers had a 1:2 ratio.

## 4. CONCLUSION

In summary, we have presented the first experimental results of a spiking element for neural networks based on coupled relaxation oscillations in superconducting nanowires. The device takes advantage of the low switching energies in nanowires to produce spikes analogous to action potentials, and is able to reproduce critical bio-realistic characteristics such as a firing threshold and refractory period. Simulations using experimentally measured device parameters enabled us to explore how nanowire somas are well-suited to specific tasks, such as a low-power inference platform, while the WTA competitions showcased how the stochastic dynamics of superconducting nanowires can be harnessed for real applications and for testing theories about behaviors observed in nature. In the future, nanowire-based WTA circuits could be combined with nanowire memory elements[23,24,25] that save the competition result as a means of establishing selectivity, while WTA subcircuits could be repeated throughout a large-scale network, like they are thought to in the brain. Energy-efficient spiking using superconducting nanowires could also be applied to event-based sensing[5] or to temporal logic architectures[26,27] that encode information in pulse arrival times.
8

**Supporting Information.**

Fabrication and design of soma, measurement details, single oscillator characterization, output signals of the input and bias ports, simulated example of a soma with uneven oscillators, and details about the image recognition and WTA simulations (PDF)

**ACKNOWLEDGEMENTS**

The authors would like to thank Mark Mondol and James Daley of the NanoStructures Lab for their technical support, Akshay Agarwal and Brenden Butters for comments on this manuscript, and Thomas Ohki and Michael Schneider for helpful discussions. This research was primarily supported by the Bose Foundation. E.T. was supported by the National Science Foundation Graduate Research Fellowship Program (NSF GRFP) under Grant No. 1122374.

**AUTHOR CONTRIBUTIONS**

E.T. designed and fabricated the devices. M. Colangelo deposited the superconducting film. E.T. designed and performed the experiments with input from K.S. and K.K.B. E.T. performed the simulations, with assistance from M. Castellani on the inductive coupling. N.L. advised the WTA simulations. E.T. wrote the manuscript with input from all of the authors. K.S. and K.K.B. supervised the project.

**REFERENCES**

1. *Rebooting the IT Revolution: A Call to Action*. https://www.semiconductors.org/resources/rebooting-the-it-revolution-a-call-to-action-2/ (2015).
2. Furber, S. Large-scale neuromorphic computing systems. *J. Neural Eng.* **13**, 051001 (2016).
3. Sourikopoulos, I. *et al.* A 4-fJ/Spike Artificial Neuron in 65 nm CMOS Technology. *Front. Neurosci.* **11**, (2017).
4. Matsumoto, R., Lequeux, S., Imamura, H. & Grollier, J. Chaos and Relaxation Oscillations in Spin-Torque Windmill Spiking Oscillators. *Phys Rev Appl.* **11**, 044093 (2019).
5. Roy, K., Jaiswal, A. & Panda, P. Towards spike-based machine intelligence with neuromorphic computing. *Nature* **575**, 607–617 (2019).
6. Bohaichuk, S. M. *et al.* Fast Spiking of a Mott $VO_2$–Carbon Nanotube Composite Device. *Nano Lett.* **19**, 6751–6755 (2019).
7. Crotty, P., Schult, D. & Segall, K. Josephson junction simulation of neurons. *Phys. Rev. E* **82**, 011914 (2010).
8. Toomey, E., Segall, K. & Berggren, K. K. Design of a Power Efficient Artificial Neuron Using Superconducting Nanowires. *Front. Neurosci.* **13**, (2019).
9. Toomey, E., Zhao, Q.-Y., McCaughan, A. N. & Berggren, K. K. Frequency Pulling and Mixing of Relaxation Oscillations in Superconducting Nanowires. *Phys. Rev. Appl.* **9**, 064021 (2018).
10. McCaughan, A. N. & Berggren, K. K. A Superconducting-Nanowire Three-Terminal Electrothermal Device. *Nano Lett.* **14**, 5748–5753 (2014).
11. Zhao, Q.-Y., McCaughan, A. N., Dane, A. E., Berggren, K. K. & Ortlepp, T. A nanocryotron comparator can connect single-flux-quantum circuits to conventional electronics. *Supercond. Sci. Technol.* **30**, 044002 (2017).
12. Ermentrout, G. B. & Terman, D. H. The Hodgkin–Huxley Equations. in *Mathematical Foundations of Neuroscience* vol. 35 1–28 (Springer New York, 2010).
13. Berggren, K. K. *et al.* A Superconducting Nanowire can be Modeled by Using SPICE. *Supercond. Sci. Technol.* (2018) doi:10.1088/1361-6668/aab149.
14. Person, R. S. & Kudina, L. P. Discharge frequency and discharge pattern of human motor units during voluntary contraction of muscle. *Electroencephalogr. Clin. Neurophysiol.* **32**, 471–483 (1972).
15. Izhikevich, E. M. Which model to use for cortical spiking neurons? *IEEE Trans. Neural Netw.* **15**, 1063–1070 (2004).
16. McCaughan, A. N., Oh, D. M. & Nam, S. W. A Stochastic SPICE Model for Superconducting Nanowire Single Photon Detectors and Other Nanowire Devices. *IEEE Trans. Appl. Supercond.* **29**, 1–4 (2019).

17. Prezioso, M. *et al.* Training and operation of an integrated neuromorphic network based on metal-oxide memristors. *Nature* **521**, 61–64 (2015).
18. Schneider, M. L. *et al.* Energy-efficient single-flux-quantum based neuromorphic computing. in *2017 IEEE International Conference on Rebooting Computing (ICRC)* 1–4 (2017). doi:10.1109/ICRC.2017.8123634.
19. Nielsen, M. Neural Networks and Deep Learning. 224.
20. Chen, Y. Mechanisms of Winner-Take-All and Group Selection in Neuronal Spiking Networks. *Front. Comput. Neurosci.* **11**, (2017).
21. Binas, J., Rutishauser, U., Indiveri, G. & Pfeiffer, M. Learning and stabilization of winner-take-all dynamics through interacting excitatory and inhibitory plasticity. *Front. Comput. Neurosci.* **8**, (2014).
22. Lynch, N., Musco, C. & Parter, M. Winner-Take-All Computation in Spiking Neural Networks. *ArXiv190412591 Cond-Mat Q-Bio* (2019).
23. Zhao, Q.-Y. *et al.* A compact superconducting nanowire memory element operated by nanowire cryotrons. *Supercond. Sci. Technol.* **31**, 035009 (2018).
24. Murphy, A., Averin, D. V. & Bezryadin, A. Nanoscale superconducting memory based on the kinetic inductance of asymmetric nanowire loops. *New J. Phys.* **19**, 063015 (2017).
25. Toomey, E. *et al.* Bridging the Gap Between Nanowires and Josephson Junctions: A Superconducting Device Based on Controlled Fluxon Transfer. *Phys. Rev. Appl.* **11**, 034006 (2019).
26. Madhavan, A., Sherwood, T. & Strukov, D. Race Logic: Abusing Hardware Race Conditions to Perform Useful Computation. *IEEE Micro* **35**, 48–57 (2015).
27. Tzimpragos, G. *et al.* A Computational Temporal Logic for Superconducting Accelerators. in *Proceedings of the Twenty-Fifth International Conference on Architectural Support for Programming Languages and Operating Systems* 435–448 (Association for Computing Machinery, 2020). doi:10.1145/3373376.3378517.

Supplemental Material

**Table of Contents**

## I. Fabrication and design

Before fabricating the soma, we considered several critical aspects of its design. First, each nanowire oscillator must have a sufficiently large series inductance such that the $L/R$ time constant is on the order of nanoseconds, allowing for relaxation oscillations. As a result, long meandered nanowire inductors (~200 squares) were added between the 60-nm-wide switching elements and the shunt resistor in order to get a series inductance on the order of nanohenries for a typical NbN film with a sheet inductance of 20–50 pH/sq. Additionally, the operation of the soma relies on the bias current splitting evenly between the two branches of the nanowire loop so that both oscillators are biased identically. To ensure that the oscillator biases were equal, COMSOL simulations of both pathways were used to check that the number of squares on either side of the bias port were the same (~634 squares).

Once the design was finalized, the soma was fabricated using a two-step electron-beam lithography process. Metal shunt resistors and alignment marks were patterned in the first process. To begin, the positive-tone resist ZEP520 was spun at 5 krpm for 60 s and baked at 180 °C for 2 min. The pattern was then exposed in a 125 kV Elionix system at a beam current of 5 nA and dose of 500 µC/cm$^2$. Following exposure, the resist was developed in o-xylene at 0 °C for 60 s and isopropyl alcohol (IPA) at room temperature for 30 s, then dried with a nitrogen gun. Afterwards, a metal layer of 10 nm Ti and 25 nm Au was evaporated and lifted off in n-methyl-2-pyrrolidone (NMP) at 60 °C for 1 hr.

The second electron-beam lithography process patterned the nanowires. A niobium nitride film with a sheet resistance of 150 Ω/square was deposited[1] on top of the metal structures. Afterwards, the positive tone resist gL2000 was spun on top at 5 krpm for 60 s and baked at 180 °C for 2 min. The nanowire structures were then exposed at a dose of 600 µC/cm$^2$, using a beam current of either 500 pA or 5 nA depending on the size of the structure. Following exposure, the patterned resist was developed in o-xylene at 5 °C for 30 s and IPA at room temperature for 30 s. After checking for pattern fidelity using a scanning electron microscope, the pattern was transferred to the underlying superconducting film using reactive ion etching in CF$_4$ at a pressure of 10 mTorr and RF power of 50 W. The resist was then stripped in heated NMP for 1 hr and left overnight.

## II. Experimental Details

For all measurements, the chip was attached to a custom printed circuit board (PCB) and immersed in liquid helium at 4.2 K. Direct electrical connections were formed by wire bonding from the superconducting leads to the gold external pads on the PCB with aluminum wire bonds. Signals were sent between the room temperature electronics and the PCB via coaxial cables with SMP connectors.

*Oscillation frequency measurements*

To measure the oscillation frequency as a function of bias current, a bias current was applied using a battery source (SRS SIM928) in series with a 100 kΩ resistor, and the output voltage was sent through a 50 dB, 1 GHz-bandwidth amplifier (MITEQ AM-1309) and read-out on an oscilloscope (LeCroy WaveRunner 620Zi). A Fast Fourier Transform of the output signal was used to identify the frequency peak.

*Input pulse measurements*

To measure spiking of the soma in response to an input pulse, we applied both a DC bias current and an input current pulse. The bias current was supplied through a DC battery source in series with a 10 kΩ resistor and a bias-tee with the RF port shorted to ground. The input pulse was generated using an Agilent waveform generator (Agilent 33600a) in series with a 100 kΩ series resistor, which was sent to both the device and the oscilloscope using a pulse splitter. The voltage output was sent through the RF port of a bias-tee and a 50 dB, 1 GHz-bandwidth amplifier (MITEQ AM-1309) before being read-out by the oscilloscope.

### III. Single oscillator characterization

To understand the spiking characteristics of the soma, it was first necessary to measure the dynamics of an isolated oscillator. Figure S1 shows the oscillation frequency as a function of bias current for an individual oscillator identical to the main and control oscillators in the nanowire soma. To measure the oscillation frequency, a bias current was applied using a battery source in series with a 100 kΩ resistor, and the output voltage was sent through a 50 dB, 1 GHz-bandwidth amplifier (MITEQ AM-1309) and read-out on an oscilloscope, as described above. A Fast Fourier Transform (FFT) of the output signal was used to identify the frequency peak.

The frequency-versus-bias curve of Fig. S1 was fit to a simplified model for nanowire relaxation oscillations[2], where the frequency is dominated by the slower time constant of the signal's falling edge:

$$\frac{1}{f} \approx \left(\frac{L}{R}\right) \ln\left(\frac{I_{bias} - I_{sw}}{I_{bias}}\right)$$

The red curve in Fig. S1 shows a fit to this expression when the switching current $I_{sw}$ = 45 µA and $L/R$ = 4.92 ns. Using the calculated number of squares in the inductor and the approximate film inductance of 30 pH/sq, we can estimate that $L \sim$ 6 nH and $R_s \sim$ 1.22 Ω.

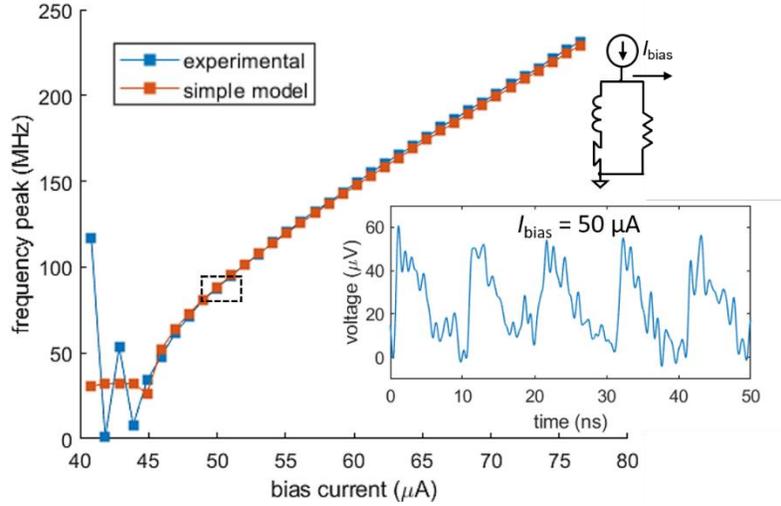

**Figure S1: Frequency of a single oscillator as a function of bias current.** Blue squares indicate experimentally measured points, while the red squares were derived from the simplified expression for oscillation frequency. The black box represents the 50 µA current bias, whose time domain characteristics are shown in the inset.

## IV. Output signals of the input and bias ports

To ensure that the spikes are coming from the phase-shifted firing of both oscillators, we examined the voltage signals of the input port and the bias port, and compared them to the simulated responses. Figure S2 (a) shows the output port (blue trace) and bias port (red trace) voltages in response to an input pulse, while the output and input port signals are shown in (b). The simulated responses for both cases are plotted in (c) and (d). By comparing Fig. S2(a) and (c), we observe that the signal from the bias port has one positive spike for each oscillator, while the traces in Fig. S2(b) and (d) show that the input port signal has one positive edge followed by one negative edge. The large spikes observed on the rising and falling edge of the input pulse in Fig. S2 (b) are likely absent from the simulation in (d), since the simulation does not account for the effects of the measurement setup, such as the amplifier and bias-tee used for readout. Despite the experimental results being noisier than the simulations, the overall agreement in the shapes of the input and bias signals indicates that the output port spikes are generated by the action of both oscillators in the loop.

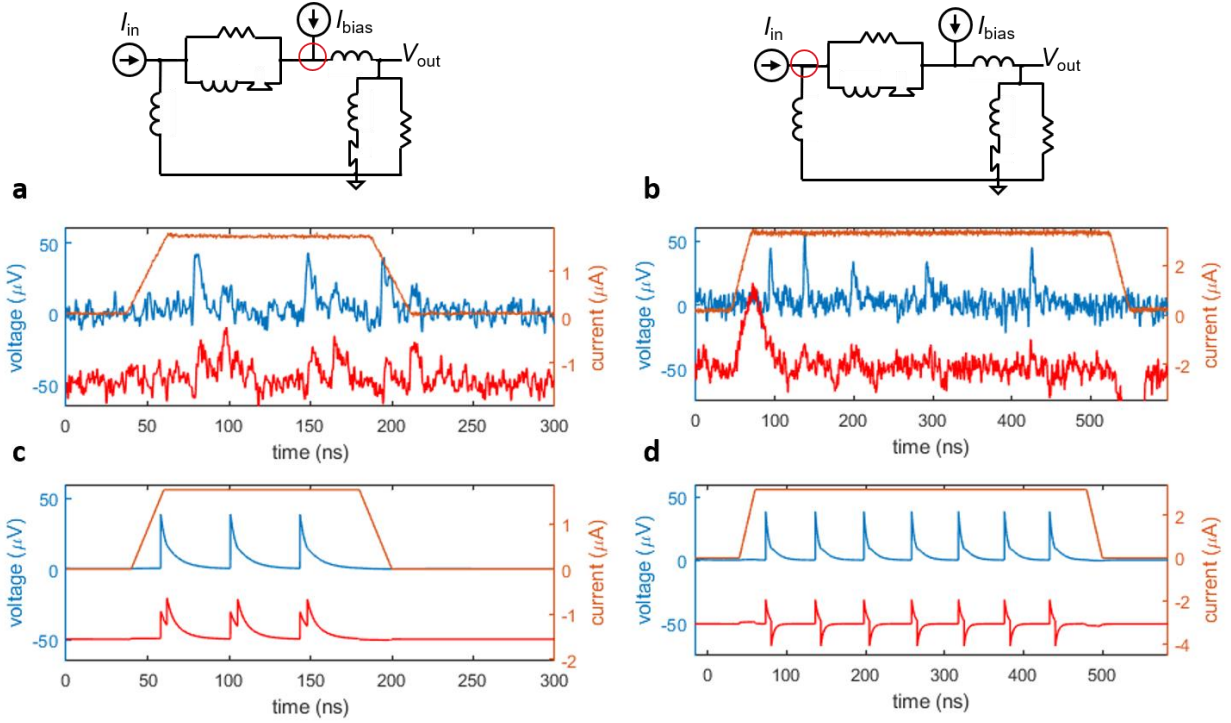

**Figure S2: Output voltages of the input and bias ports.** For all panels, the voltage from the output port is shown in blue, while the input/bias port output voltages are shown in red. Traces have been shifted on the y-axis for clarity. The orange trace shows the input current pulse. (a) Response from the bias port when $I_{bias}$ = 75.7 µA, and $I_{in}$ = 1.75 µA (pulse width = 150 ns). The bias port is indicated by the red circle in the circuit schematic above. (b) Response from the input port when $I_{bias}$ = 75.9 µA, and $I_{in}$ = 3.25 µA (pulse width = 480 ns). The input port is indicated by the red circle in the circuit schematic above. (c) Simulated response of the bias port. (d) Simulated response from the input port.

## V. Discussion on flux buildup from uneven oscillators

The firing probabilities Figure 2d of the main text showed a decrease in firing with high input currents for low bias current values. One possible explanation for this phenomenon is unequal flux or $LI_c$ products of the two oscillators that comprise the soma, where $L$ is the series inductance and $I_c$ is the critical current. If we consider the example where the control oscillator has a higher $L$, then the flux removed from the loop by the control oscillator is more than the flux added to the loop by the main oscillator. Furthermore, the time constant of the control oscillator is slower due to the higher inductance.

We can consider the example of when the loop inductors are both 12 nH and the inductance of the main oscillator is 6 nH (like in the device presented in this work), but the control oscillator inductance is 1 nH higher at 7 nH. Simulating this scenario in LTspice, we see that the control oscillator contributes roughly 115 $\Phi_0$ into loop, while the main oscillator contributes 99 $\Phi_0$ (we observed that only about ~ 33 µA of the current contributed to the $LI$ product). Since the total loop inductance is 12 nH + 12 nH + 6 nH + 7 nH = 37 nH, this translates to the main oscillator adding roughly 5.3 µA of counterclockwise circulating current, and the control oscillator removing about 6.2 µA.

The control oscillator only fires if the sum of the bias, input, and circulating currents passing through it exceeds its critical threshold. This condition may be expressed as:

$$\left| I_{bias}\left(\frac{18}{37}\right) + I_{circ} - I_{in}\left(\frac{12}{37}\right) \right| > I_c$$

where the fractions indicate the amount of bias and input currents that pass through the control oscillator by inductive splitting, and $I_{circ}$ is the amount of trapped flux stored in the form of counterclockwise circulating persistent current. As seen by this expression, the control oscillator requires more circulating current to keep firing if the input current is increased.

Figure S3a shows a calculation of the number of times the control oscillator fires before shutting off for different amounts of applied input current, based on the expression above. It assumes that the main oscillator adds 5.3 µA of current and the control oscillator removes 6.2 µA of current for every spike cycle, leading to a baseline shift of ~ -1 µA per cycle, and calculates how many times this cycle is repeated before the sum of currents through the control oscillator drops below $I_c$. Figures S3(b)-(d) show time-domain LTspice simulations when the input current is 2 µA, 4 µA, and 8 µA. As seen in the time-domain plots, the control oscillator fires after the main oscillator until the amount of circulating current is too low. For higher input currents, more circulating current is required to keep the control oscillator firing, and so it turns off after fewer cycles.

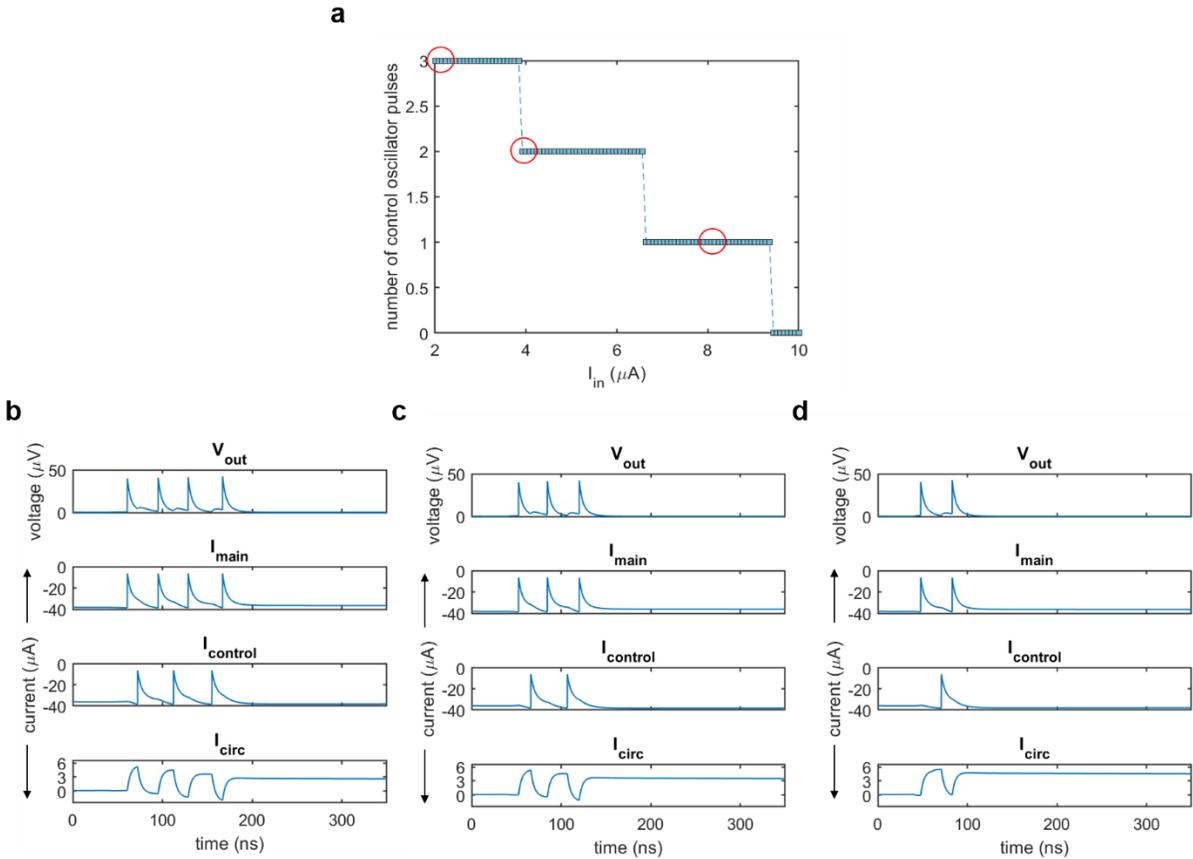

**Figure S3: Influence of uneven oscillators on soma operation.** (a) Calculation of the number of times the control oscillator is expected to fire before turning off as a function of applied input current when $I_{bias}$ = 75 µA. It was assumed that the main oscillator adds 5.3 µA of circulating current per cycle while the control oscillator removes 6.2 µA, and that total current through the control oscillator needs to be slightly higher than its $I_c$ at 38.7 µA in order to fire. Red circles indicate the input currents used in the time-domain simulations of (b)-(d). (b) LTspice simulation of the soma with uneven oscillator inductances when $I_{in}$ = 2 µA. Plots show the overall output voltage, the currents through the main and control oscillator nanowires, and the circulating current caused by the added flux. (c) Same as (b), except $I_{in}$ = 4 µA. (d) Same as (b), except $I_{in}$ = 8 µA.

The time domain simulations in Figure S3 used the deterministic soma models that have firing thresholds exactly at the defined critical current. However, as was observed in Fig. 2d of the main text, the experimental soma fires with a stochastically varying threshold. The firing probability was observed to increase with input current up to a certain point, after which it decreased with larger input currents.

We can simulate the effect of uneven oscillators on the firing probability by incorporating Gaussian noise sources into our model, like those implemented in the WTA simulations. Figure S4(a) shows a simulation of the firing probability as a function of input current, displaying a trend that is qualitatively similar to the firing rates for low bias currents of Fig. 2d in the main text. Figure S4(b) shows the respective time domain signals at two different input current points. Like in the deterministic simulations of Fig. S3, higher input currents eventually result in the control oscillator firing fewer times before shutting off. Note that the signals and spiking frequencies appear different than those of the deterministic soma in Fig. S3 due to the added noise sources.

These results indicate that uneven flux contributions from the two oscillators can lead to a nonzero baseline circulating current that eventually turns one of the oscillators off. However, as we observed experimentally, this problem may be mitigated by increasing the bias current until the firing probability curve plateaus at high input currents.

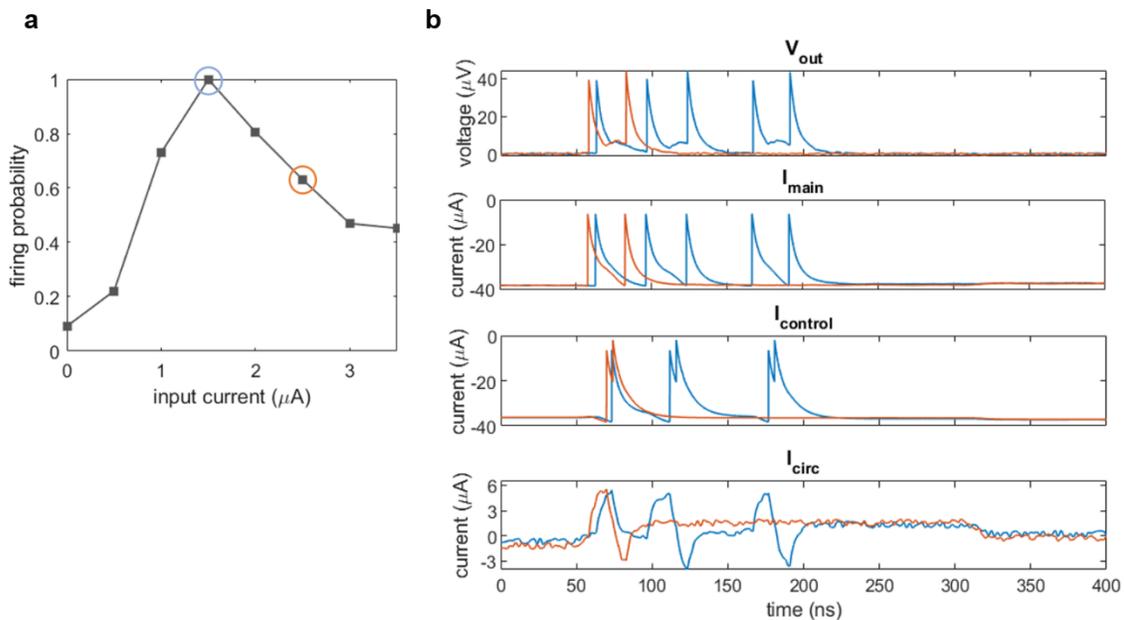

**Figure S4: Influence of uneven oscillators on simulated firing probability.** (a) Firing probability as a function of input current when $I_{bias}$ = 75.3 µA, control oscillator inductance = 7 nH and main oscillator inductance = 6 nH. Gaussian white noise sources had an amplitude of 0.8 µA. (b) Time domain of soma signals at $I_{in}$ = 1.5 µA (blue) and $I_{in}$ = 2.5 µA (red).

## VI. Pattern recognition details

Figure S5 shows the complete set of 30 images used in the image recognition circuit, based on the data set originally tested in physical memristor circuits. An external Python code was used to solve for the weight of each of synaptic connection. Table S1 shows the resulting weights when the network was trained on all 30 images. The weights were mapped onto the inductance of each synapse, with a higher weight interpreted as a lower inductance. For each synapse connecting to a pixel, the parallel resistance was 25 Ω, while the left and right resistances were 5 Ω and 6 Ω. The synapse inductance and the coupling inductance are proportional to the weight, scaled to a baseline of 0.1 μH. The magnitude of the receiving coupling inductor attached to each letter neuron was twice the synaptic inductance, and was placed in series with a 0.5 Ω resistor. Negative weights were implemented by reversing the direction of the coupling.

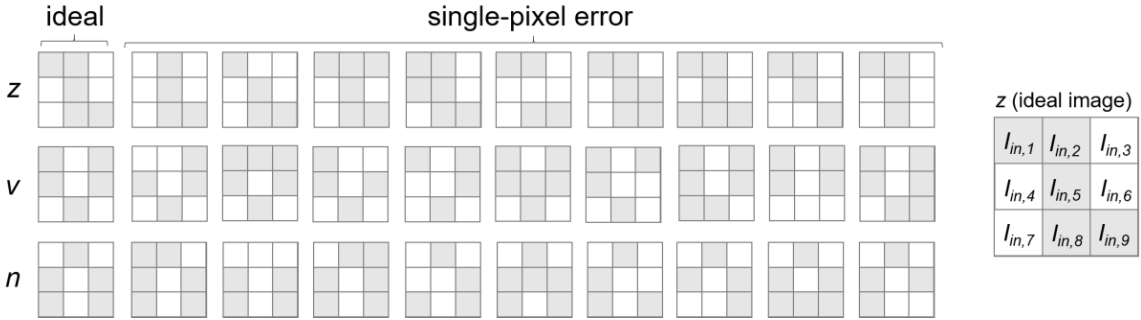

**Figure S5: Complete set of images used in the pattern recognition network.** Each letter has an ideal image and nine single-pixel error images. Pixel colors were mapped onto the input current to each pixel neuron, as indicated on the ideal z image on the righthand side. Figure adapted from M. Prezioso et al.

|   | $p_1$ | $p_2$ | $p_3$ | $p_4$ | $p_5$ | $p_6$ | $p_7$ | $p_8$ | $p_9$ |
|---|---|---|---|---|---|---|---|---|---|
| z | 1.89 | 1.04 | -0.97 | -1.85 | 0.37 | -2.33 | -1.13 | 0.80 | 1.94 |
| v | 0.31 | -1.98 | 1.98 | 0.50 | -1.14 | 0.93 | -0.80 | 1.46 | -1.14 |
| n | -1.53 | 0.99 | -1.35 | 1.66 | -0.81 | 0.14 | 2.23 | -1.71 | 0.40 |

**Table S1: Synaptic weights of the pattern recognition network using all 30 images for training.** Weights are converted into inductances for each synapse. A higher weight is translated into a lower inductance, leading to more synaptic current.

In the case presented in the main text, we used all 30 images for both training and testing because the data set is small, leading to 100% classification. However, for larger data sets, we would like to be able to solve for the synaptic weights using only a fraction of the total images as a training set, and then test the resulting network on all of the images. To show how our network would respond in this scenario, we repeated the training process using just 9 images that were randomly chosen from the set of 30, and then modified our circuit with these new synaptic weights. The results shown in Figure S6 indicate that the circuit correctly identifies 23 of the 30 images in the complete set, which is to be expected in comparison to the previous results since not all of the images were used during training. We found that the classification could be improved by non-randomly selecting 12 training images (one ideal image and three single-pixel error images, per letter), leading to correct identification of 27 of the 30 images. Given that our circuit design has not undergone any optimization, we believe that our classification results could be further improved by refining our parameters through processes like Bayesian optimization.

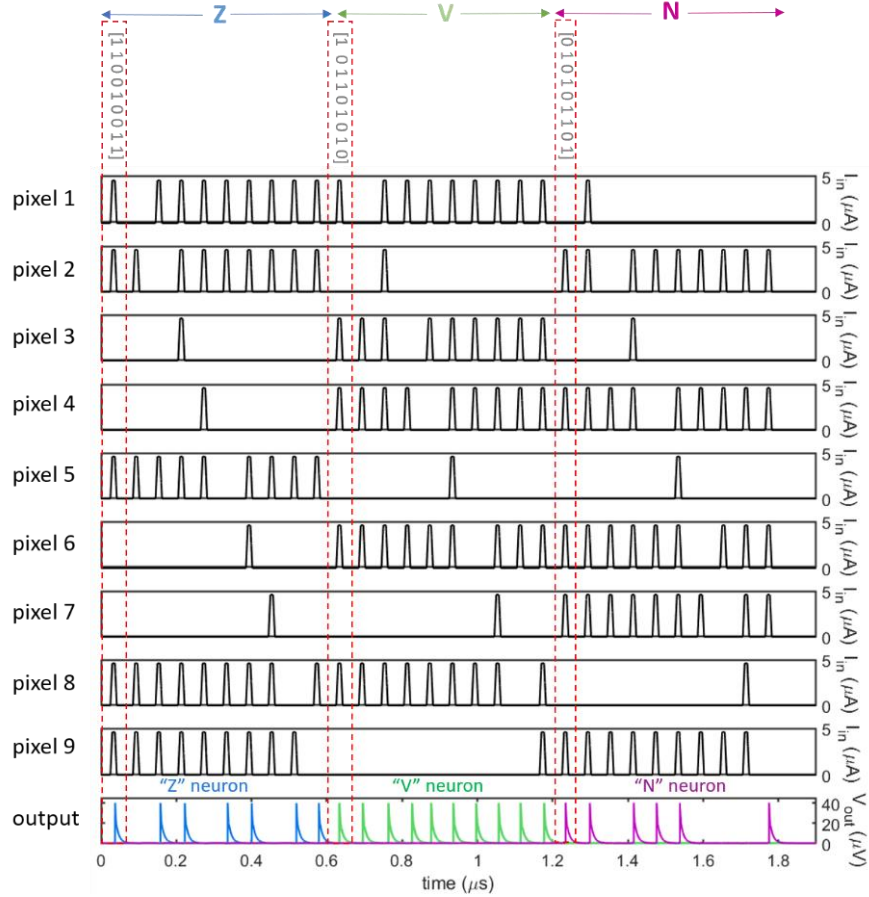

**Figure S6: Simulation of each of the 30 test images using a training set of 9 randomly chosen images.** 23 of the 30 images were correctly identified. The letter neurons were biased at 76.9 μA and had input currents of 5.45 μA.

## VII. WTA simulation details

Figure S7 shows simplified circuit schematics for each of the components in the WTA simulation. Each input neuron excites its own output neuron through positive inductive coupling, as shown in (a). Fig. S7(b) displays one of the two inhibitory neurons, which receive excitatory signals from all three output neurons, and suppress each of the output neurons through negative inductive coupling on its output. Fig. S7(c) illustrates one of the three output neurons. It receives two excitatory inputs from its particular input neuron and from its own output, and receives two inhibitory inputs from the two inhibitors.

     As noted in Figure 5a of the main text, the connections between neurons are weighted according to the dimensionless parameter $\gamma$—X:Y connections have strength $3\gamma$, Y:Y connections have strength $2\gamma$, and Z:Y and Y:Z have connections of $-\gamma$ and $\gamma$, respectively. In our system, weight corresponds to inductance, with higher weights interpreted as lower inductances, leading to more synaptic current. In the WTA circuit, all of the synapse inductances were scaled relative to 0.77 μH. For the simulations shown in Fig. 5 of the main text, the parallel synapse resistors were 15 Ω, the series resistors on the synapse outputs were both 3 Ω, and the series resistors on the inputs were 0.2 Ω. Inhibitory connections were performed through negative inductive coupling, and the couplers had a ratio of 1:2. Input neurons were biased at 77 μA and received input currents of 4 μA. The control inhibitor was biased at 76.9 μA and the stability

inhibitor was biased at 76.97 µA, with both receiving input currents of 0.8 µA. Output neurons were biased at 76.4 µA with input currents of 1.55 µA.

The Gaussian white noise sources at the input and bias ports of the output neurons had a bandwidth of 1 GHz and amplitude of 0.8 µA. To simulate the effects of probability, the noise signal of each output neuron was shifted in time by a different factor, giving them each effectively a different chance of switching (this method was chosen since it was not possible to randomize the seed at the start of successive simulations). For each competition, three unique random numbers were generated in Matlab and used as the time shift factors for each of the three output neurons.

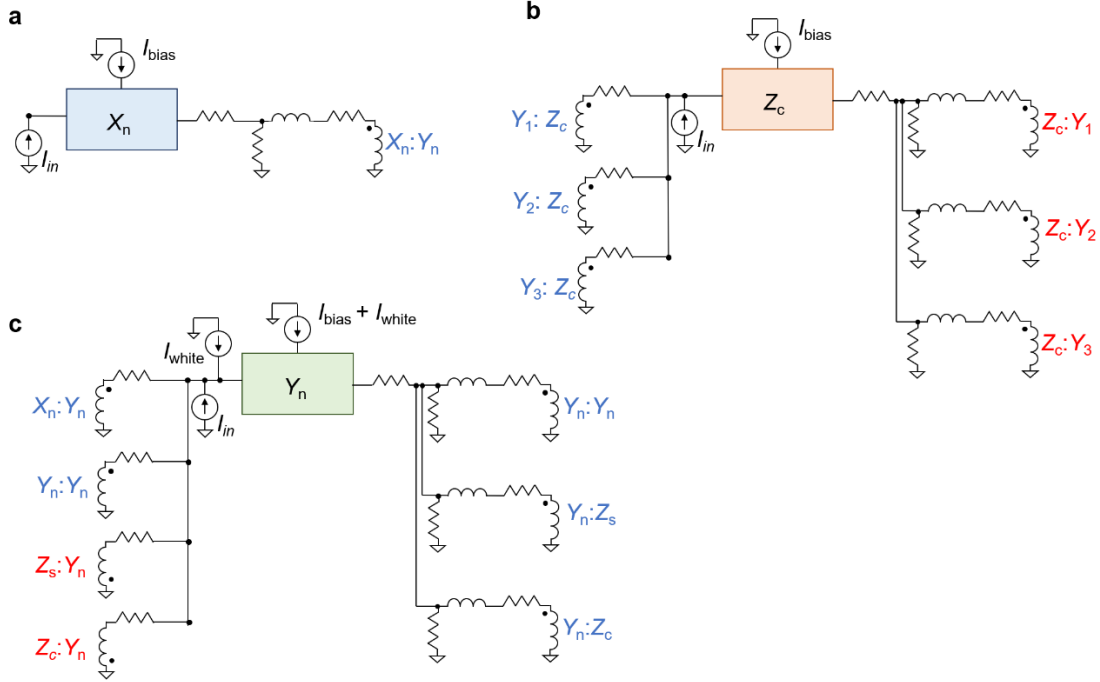

**Figure S7: Simplified circuit schematics of the WTA components.** (a) Example of an input neuron. (b) Example of one of the two inhibitors. Each output neuron $Y$ excites the inhibitor through positive inductive coupling. The output of the inhibitor is negatively coupled to each of the output neurons in order to suppress them. (c) Example of an output neuron. Each output neuron is excited by an input neuron and its own output, and is suppressed by both inhibitors. The output is split into three connections to excite both inhibitors and the output neuron itself. Gaussian white noise sources at the input and bias ports are used to create firing probabilities that match what is observed experimentally. For all diagrams, excitatory coupling is indicated by blue text, and inhibitory coupling is indicated by red text.

**References:**

1. Dane, A. E. *et al.* Bias sputtered NbN and superconducting nanowire devices. *Appl. Phys. Lett.* **111**, 122601 (2017).
2. Toomey, E., Zhao, Q.-Y., McCaughan, A. N. & Berggren, K. K. Frequency Pulling and Mixing of Relaxation Oscillations in Superconducting Nanowires. *Phys. Rev. Appl.* **9**, 064021 (2018).